\documentclass[manuscript,trackchanges]{aastex62}
\usepackage{amsmath}

\shorttitle{pyCallisto: A Python Library To Process The CALLISTO Spectrometer Data}
\shortauthors{Pawase et al.}

\usepackage{url}
\usepackage{csquotes}
\usepackage{graphicx}
\usepackage{listings}
\usepackage{color}
\usepackage{hyperref}
\usepackage[normalem]{ulem}

\definecolor{codegreen}{rgb}{0,0.6,0}
\definecolor{codegray}{rgb}{0.5,0.5,0.5}
\definecolor{codepurple}{rgb}{0.58,0,0.82}
\definecolor{backcolour}{rgb}{0.95,0.95,0.92}
 
\lstdefinestyle{mystyle}{
    backgroundcolor=\color{backcolour},   
    commentstyle=\color{codegreen},
    keywordstyle=\color{magenta},
    numberstyle=\tiny\color{codegray},
    stringstyle=\color{codepurple},
    basicstyle=\footnotesize,
    breakatwhitespace=false,         
    breaklines=true,                 
    captionpos=b,                    
    keepspaces=true,                 
    numbers=left,                    
    numbersep=5pt,                  
    showspaces=false,                
    showstringspaces=false,
    showtabs=false,                  
    tabsize=2
}
 
\begin{document}

\title{pyCallisto: A Python Library To Process The CALLISTO Spectrometer Data}

\correspondingauthor{Ravindra Pawase}
\email{ravi.pawase@gmail.com}

\author{Ravindra Pawase}
\affiliation{Balewadi, Pune - 411 045, India.} 

\author[0000-0002-1192-1804]{K. Sasikumar Raja}

\affiliation{Indian Institute of Science Education and Research, Pashan, Pune - 411 008, India.} 






\begin{abstract}
CALLISTO is a radio spectrometer designed to monitor the transient radio emissions / bursts originated from the solar corona in the frequency range $45-870$ MHz. At present there are 
$\gtrsim 150$ stations (together forms an e-CALLISTO network) around the globe continuously monitoring the Sun 24 hours a day. We have developed a \emph{pyCallisto}, a python library to process the CALLISTO data observed by all stations of the e-CALLISTO network. In this article, we demonstrate various useful functions that are routinely used to process the CALLISTO data with suitable examples. This library is not only efficient in processing the data but plays a significant role in developing automatic classification algorithms of different types of solar radio bursts. \\

\end{abstract}


\section{Introduction}

A Compound Astronomical Low cost Low frequency Instrument for Spectroscopy and Transportable Observatory (CALLISTO), is a radio spectrometer to monitor the transient radio emissions / bursts from the solar corona \citep{Ben2005, Ben2009}. The CALLISTO operates from 45 to 870 MHz. There are $\gtrsim 150$ stations distributed around the world and all together forms an e-CALLISTO network (\url{http://www.e-callisto.org/}). 
As the spectrometers are distributed over different longitudes, we can monitor the radio emissions 24 hours a day. Each station stores one data file (in FITS format) in every 15 minutes and fetches to the server located at ETH Zurich, Switzerland. Note that different stations operate over different bandwidths depending on the radio frequency interference (RFI) and the instrumentation limitations. On average each station carry out the observations for 9 hours a day. The real time data is made accessible to the public via e-CALLISTO web-page (\url{http://soleil.i4ds.ch/solarradio/callistoQuicklooks/}). 

In order to process the data, we have developed a python library called \emph{pyCallisto} with routinely used subroutines/functions. This library source code is made available in the public domain for users via git-hub (\url{https://github.com/ravipawase/pyCallisto}). In this article, we describe various functionalities that are useful in processing the data and provide a step by step guide to the users. We expect that this article will be helpful for the users. 

\section{Installing anaconda and dependencies}

In order to process the data using the \emph{pyCallisto} library, we recommend to install the open source anaconda distribution with Python version 3 or above. Note that this library may not work for the python versions below 3. On top of it, we need to install the following standard python packages: \emph{Numpy} \citep{numpy}, \emph{Matplotlib} \citep{matplotlib}, \emph{Astropy} \citep{astropy}. Note that Anaconda python distribution is available for Windows, OS X and Linux operating systems (\url{https://www.anaconda.com/distribution/}) and this library works efficiently over all the operating systems.  
We note here that this library was thoroughly tested in Ubuntu 16.04 LTS operating system. However we do not see any reason for not working in other operating systems.

\section{Features of \emph{pyCallisto} library}\label{sec:list}

\noindent Various functionalities developed under \emph{pyCallisto} library are described here. In this article we use the data observed using the CALLISTO spectrometer located at Indian Institute of Science Education and Research (IISER), Pune, India (longitude $73^{\circ}~ 55'$ E,  latitude $18^{\circ}~ 31'$ N and situated at an altitude of 558 meters above the sea level) on 2015 November 04  \citep{Sas2018}. The basic functionalities of the library are the following:

\begin{itemize}
\item \emph{spectrogram}: plots a dynamic spectrogram of the fits file
\item \emph{appendTimeAxis}: joins two spectrograms along the time axis
\item \emph{sliceTimeAxis}: crops desired part of spectrogram along time axis
\item \emph{sliceFrequencyAxis}: crops desired part of spectrogram along frequency axis
\item \emph{subtractBackground}: estimates and subtracts the background from the spectrogram
\item \emph{meanLightCurve}: generates a mean light curve averaged over frequency axis (i.e., time vs amplitude / intensity) 
\item \emph{meanSpectrum}: generates a mean spectrum averaged over time axis (i.e., frequency vs amplitude / intensity)
\item \emph{lightCurve}: generates the light curve at a given frequency channel
\item \emph{spectrum}: generates the spectrum at a given time sample
\item \emph{universalPlot}: plots the spectrogram, mean light curve and mean spectrum together. 
\end{itemize}

\noindent Firstly, \emph{pyCallisto} library has be downloaded from git-hub page \url{https://github.com/ravipawase/pyCallisto}. Then we need to import the standard python libraries like \emph{numpy}, \emph{matplotlib} and \emph{astropy}. The details of importing \emph{pyCallisto} library and the details of above mentioned functionalities are described in the following sub-sections. 

\subsection{Importing \emph{pyCallisto}}

\noindent At present, we recommend to keep a copy of the files \emph{pyCallisto.py} and \emph{pyCallistoUtils.py} in the current working directory (i.e., the directory where your main program is located). Otherwise, we have to set the path of the directory where these programs are located using the command shown in Listing \ref{l:cmd}.

\begin{lstlisting}[language=Python, label={l:cmd}, caption=Basic imports and data files]
sys.path.append('/path/to/pyCallisto/filename')
\end{lstlisting}

In the following example (see Listing \ref{l:imp}), copy of the files \emph{pyCallisto.py} and \emph{pyCallistoUtils.py} are kept in ``src" folder which is inside the parent folder of the current script. 
\begin{lstlisting}[language=Python, label={l:imp}, caption=Basic imports and data files]
import sys
sys.path.append('../src/') 
import pyCallisto as pyc
import pyCallistoUtils as utils

import astropy.io.fits as pyfits
import matplotlib.pyplot as plt

#fits data files path
fits1_path = '../data/IISERP_20151104_031152_59.fit'
fits2_path = '../data/IISERP_20151104_032652_59.fit'
\end{lstlisting}

\noindent First, one has to create a \emph{pyCallisto} object which can be used further on functions listed in Section \ref{sec:list} whenever it is required (see Listing \ref{l:obj}).

\begin{lstlisting}[language=Python, label={l:obj}, caption=Creates pyCallisto object]
fits1 = pyc.pyCallisto.fromFile(fits1_path)
print("Type of the fits1 is %s"%type(fits1))
\end{lstlisting}

\subsection{Spectrogram}
\begin{figure}[!ht]
\centering
  \includegraphics[scale=0.5]{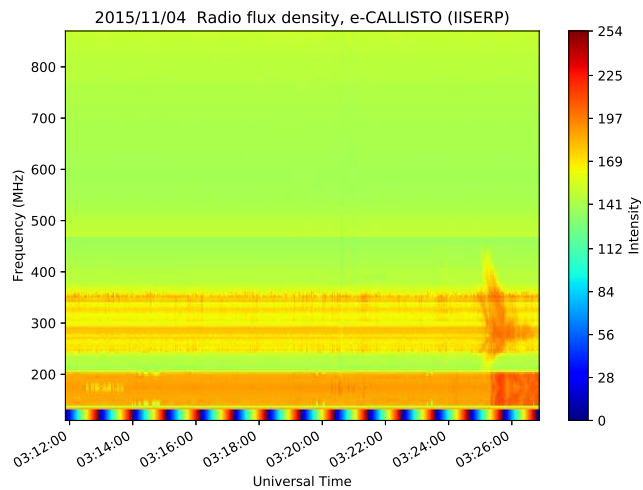}
  \caption{Spectrogram of File-1}
  \label{fig:plot1}
\end{figure}

The \emph{spectrogram} function plots the spectrogram by making use of the object that we have created in Listing \ref{l:obj}. We have given the optional inputs like: \emph{xtick} which help in deciding required number of ticks in x axis (in mins) and \emph{blevel} which decides on background level. This function returns a \emph{matplotlib} \emph{plt} object which can be further used if the user needs to save the image. The code shown in Listing \ref{l:plot1} plots the spectrogram shown in Figure \ref{fig:plot1}\\

\begin{lstlisting}[language=Python, label={l:plot1}, caption= Plots a spectrogram]
plot1 = fits1.spectrogram()
\end{lstlisting}

\begin{figure}[!ht]
\centering
  \includegraphics[scale=0.5]{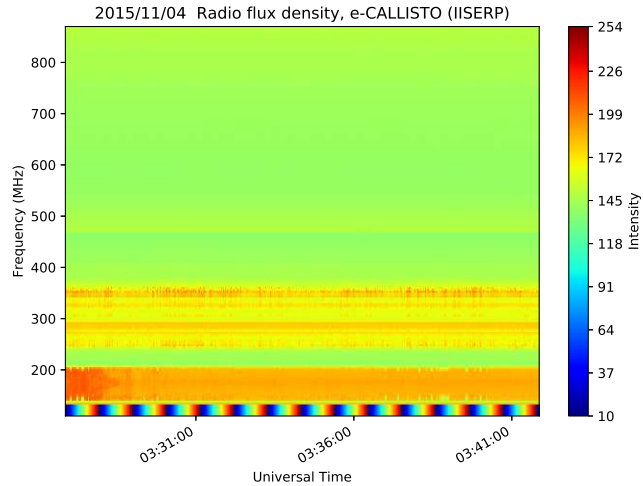}
  \caption{Spectrogram from File-2}
  \label{fig:plot2}
\end{figure}

\noindent We plot another spectrogram using Listing - \ref{l:plot2} by pass \emph{xticks} = 5 and \emph{blevel}=10. The corresponding spectrogram is shown in Figure \ref{fig:plot2}\\

\begin{lstlisting}[language=Python, label={l:plot2}, caption= Plots spectrogram-2]
fits2 = pyc.pyCallisto.fromFile(fits2_path)
plot2 = fits2.spectrogram(xtick=5, blevel=10)
\end{lstlisting}

\noindent The Figures \ref{fig:plot1} and \ref{fig:plot2} can be saved in any format (for example, png, eps , jpeg etc) using the Listing - \ref{l:save}. 

\begin{lstlisting}[language=Python, label={l:save}, caption=Saves the image returned by the spectrogram function]
plot2.savefig("first_plot.png")
\end{lstlisting}

\subsection{appendTimeAxis}

\begin{figure}[!ht]
\centering
  \includegraphics[scale=0.5]{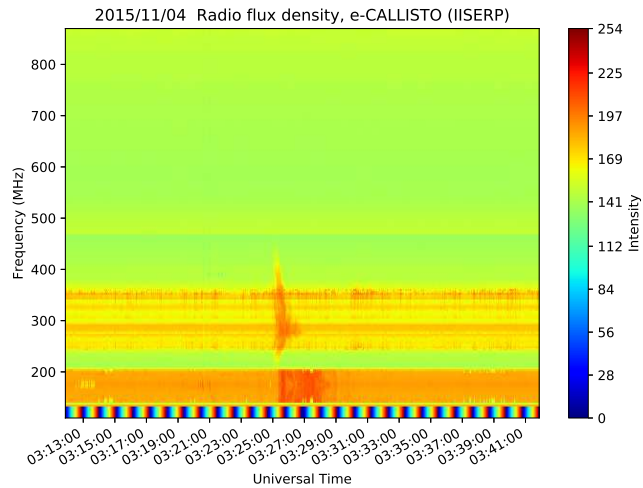}
  \caption{Combined spectrogram}
  \label{fig:join}
\end{figure}

The \emph{appendTimeAxis} function is used to combine two spectrograms along the time axis. We make a note here that the input spectrograms should be continuous in time. (Refer Listing \ref{l:combine} and Figure \ref{fig:join}). \\
\begin{lstlisting}[language=Python, label={l:combine}, 
caption= Combines the spectrogram-1 and 2]
joined = fits1.appendTimeAxis(fits2_path)
plt = joined.spectrogram()
plt.savefig("joined.png")
\end{lstlisting}

\subsection{sliceTimeAxis}
\begin{figure}[!ht]
\centering
  \includegraphics[scale=0.5]{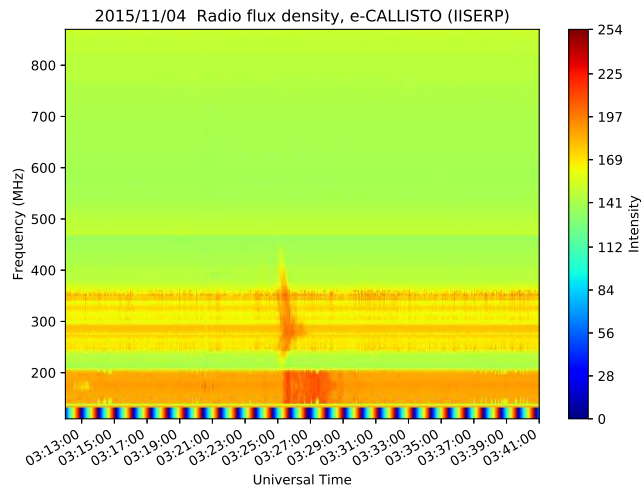}
  \caption{The spectrogram after slicing over the required time interval}
  \label{fig:tslice}
\end{figure}

\noindent The \emph{sliceTimeAxis} takes two inputs, start time and end time; it returns the spectrogram within the given time limits (Refer Listing \ref{l:slicetime} and Figure \ref{fig:tslice}).\\

\begin{lstlisting}[language=Python, label={l:slicetime}, caption= Slices the spectrogram along time axis]
time_sliced = joined.sliceTimeAxis("03:12:00", "03:41:00")
plt = time_sliced.spectrogram()
plt.savefig("time_sliced.png")
\end{lstlisting}

\subsection{sliceFrequencyAxis}
\begin{figure}[!ht]
\centering
  \includegraphics[scale=0.5]{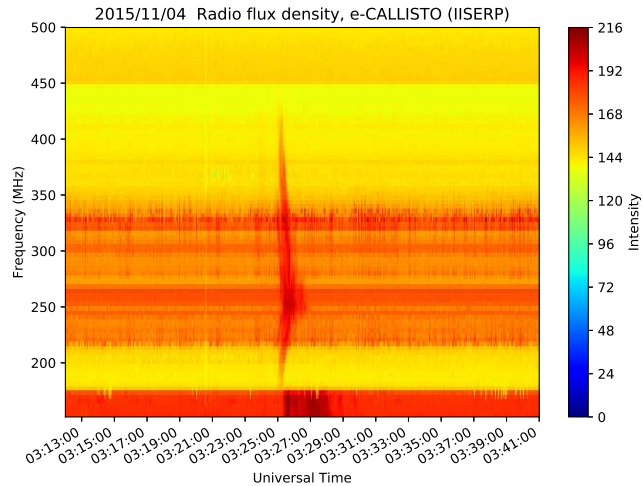}
  \caption{The spectrogram after slicing over the required frequency range}
  \label{fig:fslice}
\end{figure}

\noindent The \emph{sliceFrequencyAxis} function takes two inputs, start frequency and end frequency. This function slices the spectrogram within the frequency limits that are provided
(Refer Listing \ref{l:slicefreq} and Figure \ref{fig:fslice}).\\

\begin{lstlisting}[language=Python, label={l:slicefreq}, caption= Slices the spectrogram along frequency axis]
freq_sliced = time_sliced.sliceFrequencyAxis(150, 500)
plt = freq_sliced.spectrogram()
plt.savefig('frequency_sliced.png')
\end{lstlisting}

\subsection{subtractBackground}
\begin{figure}[!ht]
\centering
  \includegraphics[scale=0.5]{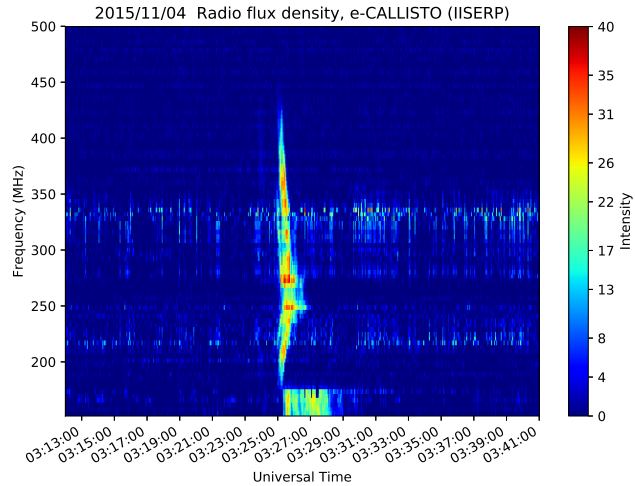}
  \caption{The spectrogram after the background subtraction}
  \label{fig:sbg}
\end{figure}

\noindent The \emph{subtractBackground} does not take any input but works on the \emph{pycallisto} object. This function calculates the median of each frequency channel and subtracts it from corresponding channel (Refer Listing \ref{l:sbg} and Figure \ref{fig:sbg}).

\begin{lstlisting}[language=Python, label={l:sbg}, caption={Subtracts the background from the spectrogram}]
background_subtracted = freq_sliced.subtractBackground()
plt = background_subtracted.spectrogram()
plt.savefig("background_subtracted.png")
\end{lstlisting}

\subsection{meanLightCurve}
\begin{figure}[!ht]
\centering

  \includegraphics[scale=0.5]{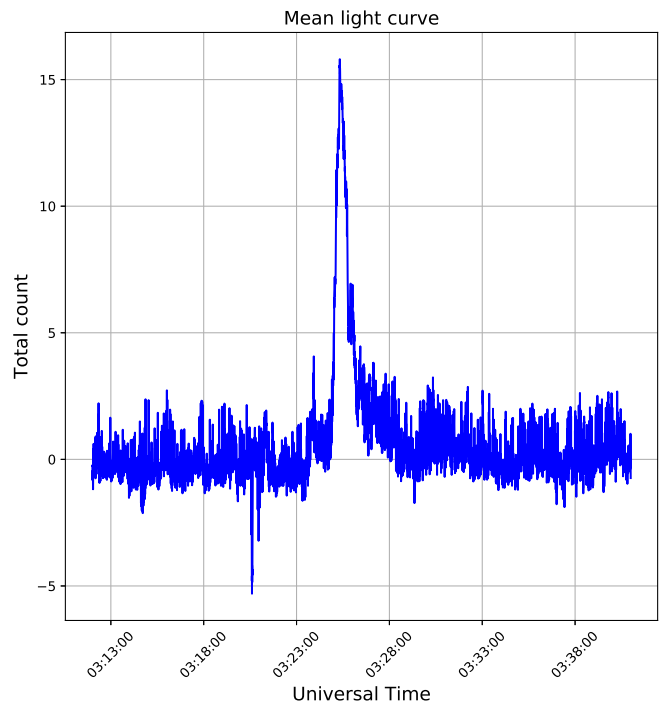}
  \caption{The light curve averaged over all frequencies}
  \label{fig:mlc}
\end{figure}

\noindent The \emph{meanLightCurve} takes two inputs, out file name and grid which is a Boolean parameter that provides an option to plot the grid or not. This generates the light curve averaged over all frequencies of the spectrogram (Refer Listing \ref{l:mlc} and Figure \ref{fig:mlc}).\\

\begin{lstlisting}[language=Python, label={l:mlc}, caption= Creates the mean light-curve]
background_subtracted.meanLightCurve(outImage ="mean_Light_Curve.png", grid=True)
\end{lstlisting}

\subsection{meanSpectrum}
\begin{figure}[!ht]
\centering
  \includegraphics[scale=0.5]{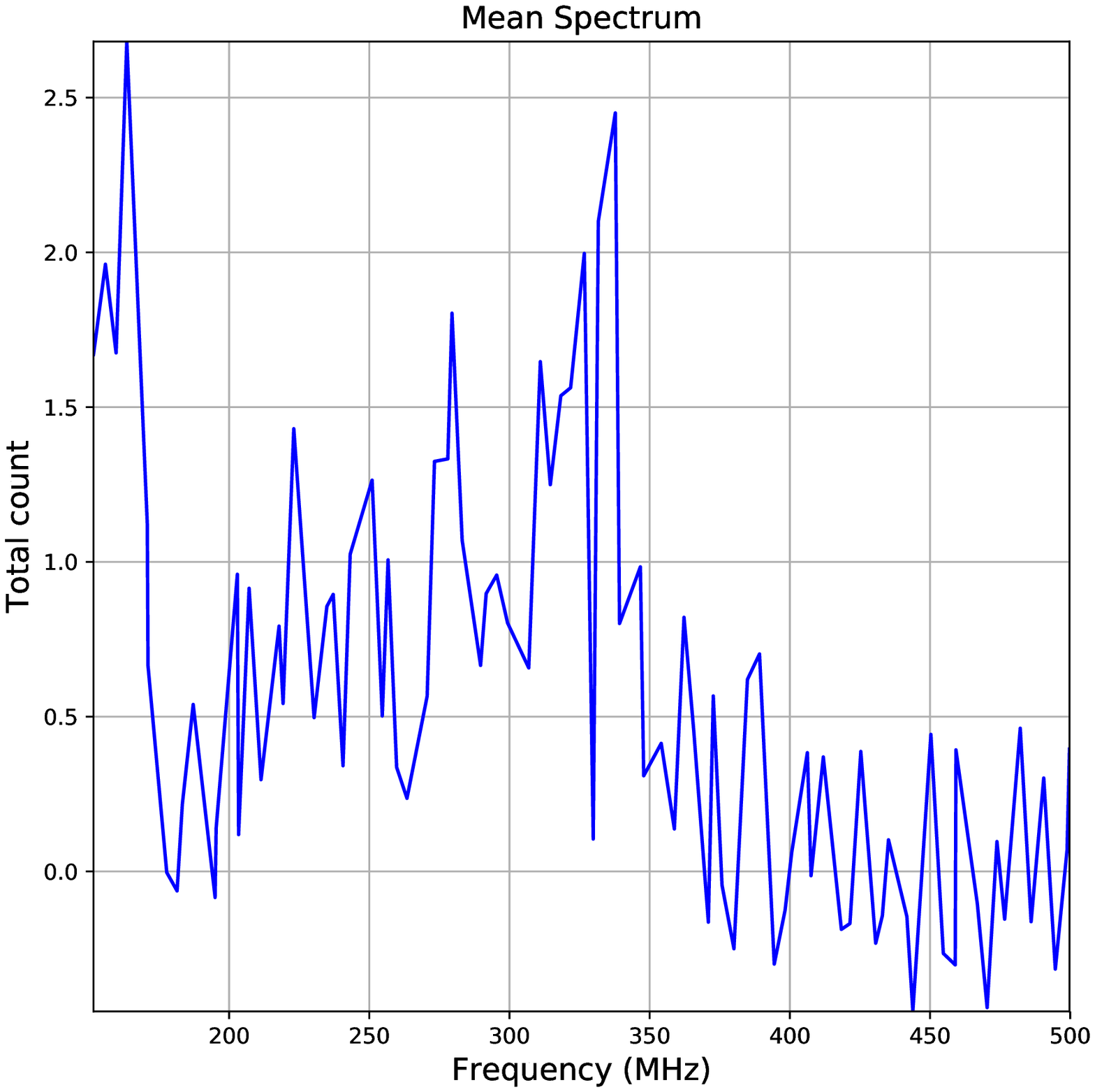}
  \caption{The spectrum averaged over time axis}
  \label{fig:mspectrum}
\end{figure}

\noindent The \emph{meanSpectrum} takes two inputs: out file name and grid which is a Boolean operator and that decides to keep the grid on or off. This function plots the spectrum by averaging over the time axis (Refer Listing \ref{l:meanspec} and Figure \ref{fig:mspectrum}).\\

\begin{lstlisting}[language=Python, label={l:meanspec}, caption= Creates mean spectrum]
background_subtracted.meanSpectrum(outImage ="mean_spectrum.png", grid=True)
\end{lstlisting}

\subsection{lightCurve}
\begin{figure}[!ht]
\centering
  \includegraphics[scale=0.5]{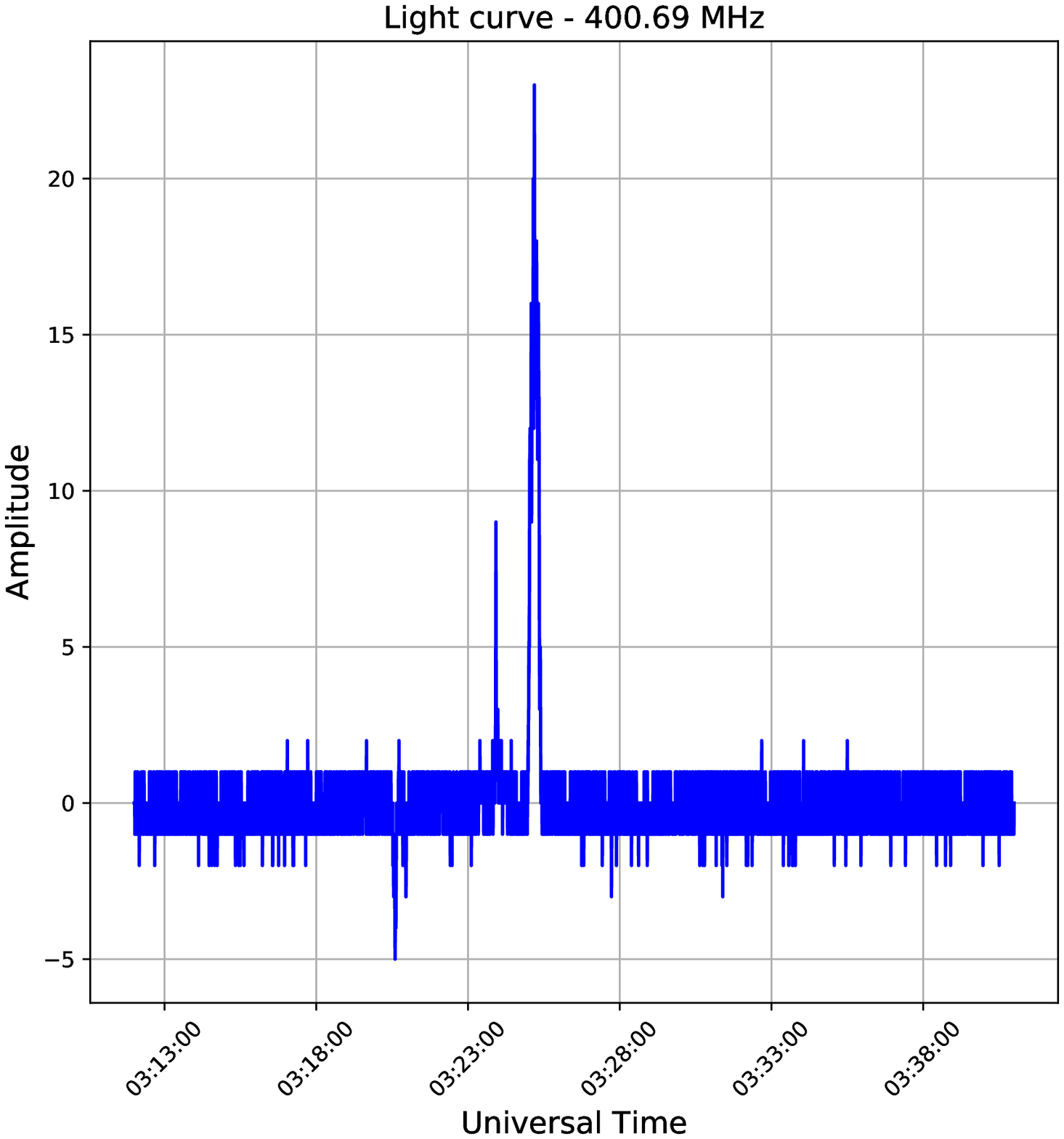}
  \caption{Light curve at a selected frequency}
  \label{fig:specific_lc}
\end{figure}

\noindent The \emph{lightcurve} generates a simple light curve at selected frequency.
It takes three inputs, the frequency at which we need to plot light curve, out file name and \emph{grid} which is a Boolean value that decides to plot grid or not
(Refer Listing \ref{l:lc} and Figure \ref{fig:specific_lc}).\\

\begin{lstlisting}[language=Python, label={l:lc}, caption= Creates light curve at a selected frequency]
background_subtracted.lightCurve(400, outImage ="Lightcurve1.png", grid=True)
\end{lstlisting}

\subsection{spectrum}
\begin{figure}[!ht]
\centering
  \includegraphics[scale=0.5]{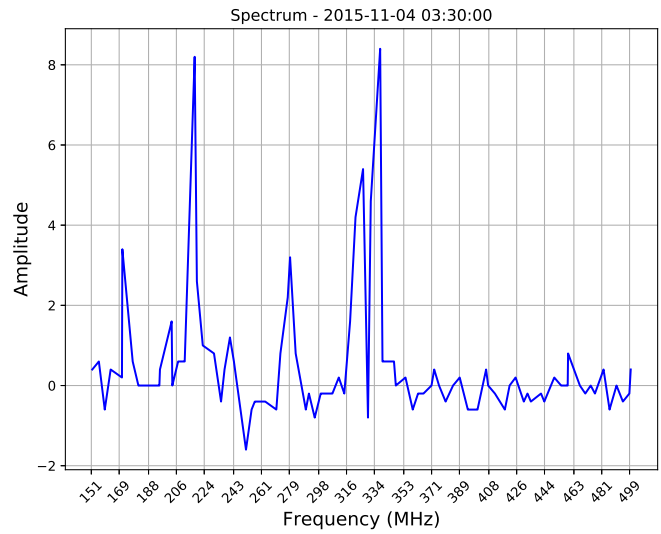}
  \caption{Spectrum at a selected time}
  \label{fig:specific_spectrum}
\end{figure}

\noindent The \emph{spectrum} generates a spectrum at given time.
It takes four inputs, date, time at which we need to plot a spectrum, out file name and and the grid which is Boolean value to turn on or off the grid (Refer Listing \ref{l:spectrum} and Figure \ref{fig:specific_spectrum}).\\

\begin{lstlisting}[language=Python, label={l:spectrum}, caption= Creates spectrum at selected time]
background_subtracted.spectrum( '2015/11/04','03:30:00', outImage ="singletimespectrum.png", grid=True)
\end{lstlisting}

\subsection{universalPlot}
\begin{figure}[!ht]
\centering
  \includegraphics[scale=0.7]{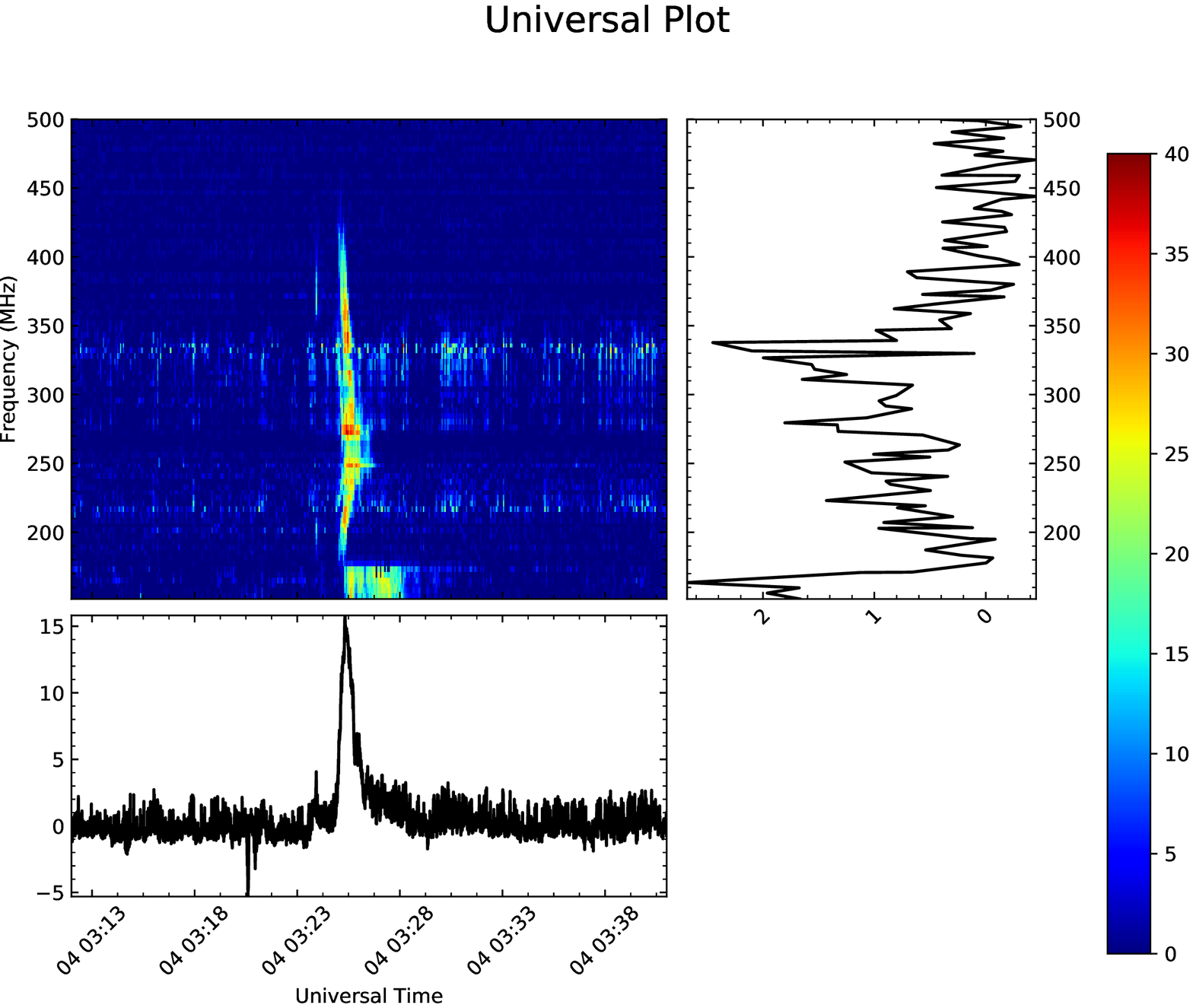}
  \caption{The Figure shows the spectrogram, mean light curve and mean spectrum together.}
  \label{fig:univ}
\end{figure}

\indent The \emph{universalPlot} plots a spectrogram along with mean spectrum and mean light curve together. It takes two inputs, out file name and title of the plot 
(Refer Listing \ref{l:univ} and Figure \ref{fig:univ}).\\

\begin{lstlisting}[language=Python, label={l:univ},  caption= {Plots spectrogram, mean light curve and mean spectrum of the spectrogram)}]
background_subtracted.universalPlot(plotName ="universal_plot_with_add_processing.png", title='Universal Plot')
\end{lstlisting}

\section{Summary and Conclusions}

\noindent We have developed a \emph{pyCallisto} python library to plot and process the data obtained using CALLISTO spectrometers (of e-CALLISTO network) that are located at different longitudes around the globe to monitor the radio transient emissions from the solar corona. In the article we have described the \emph{pyCallisto} library and various routinely used functionalities developed by us with suitable examples. In this article, we have used the data observed using CALLISTO spectrometer located at IISER, Pune, India. This library is efficient in analysing the data obtained by all stations as the data formats are more or less same. We believe that this small piece of library reduces the effort of every beginner to develop their own data analysis programs. Further this library will play a significant role in developing automatic classification algorithms of different types of solar radio bursts \citep[\emph{e.g.} see][]{Day2019}.

\section*{Acknowledgment}
K.S.R. acknowledges the financial support from the Science \& Engineering Research Board (SERB), Department of Science \& Technology, India (PDF/2015/000393). 
\bibliographystyle{aasjournal}
\bibliography{ms}

\end{document}